\documentclass[a4paper,11pt]{article}
\usepackage{pos}
\usepackage[capitalise]{cleveref}

\title{On Consistent Kinetic Mixing and the Higgs Low-Energy Theorems}
\ShortTitle{Consistent Kinetic Mixing}
\author*{Patrick Foldenauer}
\affiliation{Instituto de F\' isica Te\'orica UAM/CSIC,\\ Universidad Aut\'onoma de Madrid, 28049 Madrid, Spain}
\emailAdd{patrick.foldenauer@csic.es}
\abstract{
A popular class of extensions of the Standard Model (SM) are models of a new Abelian gauge boson $X$, called \textit{dark} or \textit{hidden photon}, that kinetically mixes with the SM photon. 
We revisit the matching procedure of kinetic mixing terms in the electroweak symmetric phase to the ones in the broken phase. Our central finding is that in order to obtain the correct matching prescription one has to take into account mixing of the hidden photon with the neutral component of the weak $SU(2)_L$ bosons. This mixing is generated by a dimension-six operator and, in theories where $SU(2)_L$ multiplets are charged under the novel Abelian gauge group, is necessarily induced at the one-loop level. We illustrate this matching procedure for the loop-generated kinetic mixing in $U(1)_{L_\mu-L_\tau}$. Furthermore, we show how to obtain general expressions for the Higgs decay amplitudes to two neutral vector bosons from the vacuum polarisation amplitudes  via the low-energy theorems. As an application, we derive general expression for the branching ratios of the decays $h\to\gamma X$ and $h\to XX$ in $U(1)_{B-L}$.
}
\FullConference{%
  8th Symposium on Prospects in the Physics of Discrete Symmetries (DISCRETE 2022)\\
  7-11 November, 2022\\
  Baden-Baden, Germany
}

\begin{document}
\maketitle

%%%%%%%%%%%%%%%%%%%%%%%%%%%%%%%%%%%%%%%%%%%%%%%
\section{Introduction}
\label{sec:introduction}
%%%%%%%%%%%%%%%%%%%%%%%%%%%%%%%%%%%%%%%%%%%%%%%

\begin{figure}[b]
    \centering
    \includegraphics[width=.9\textwidth] {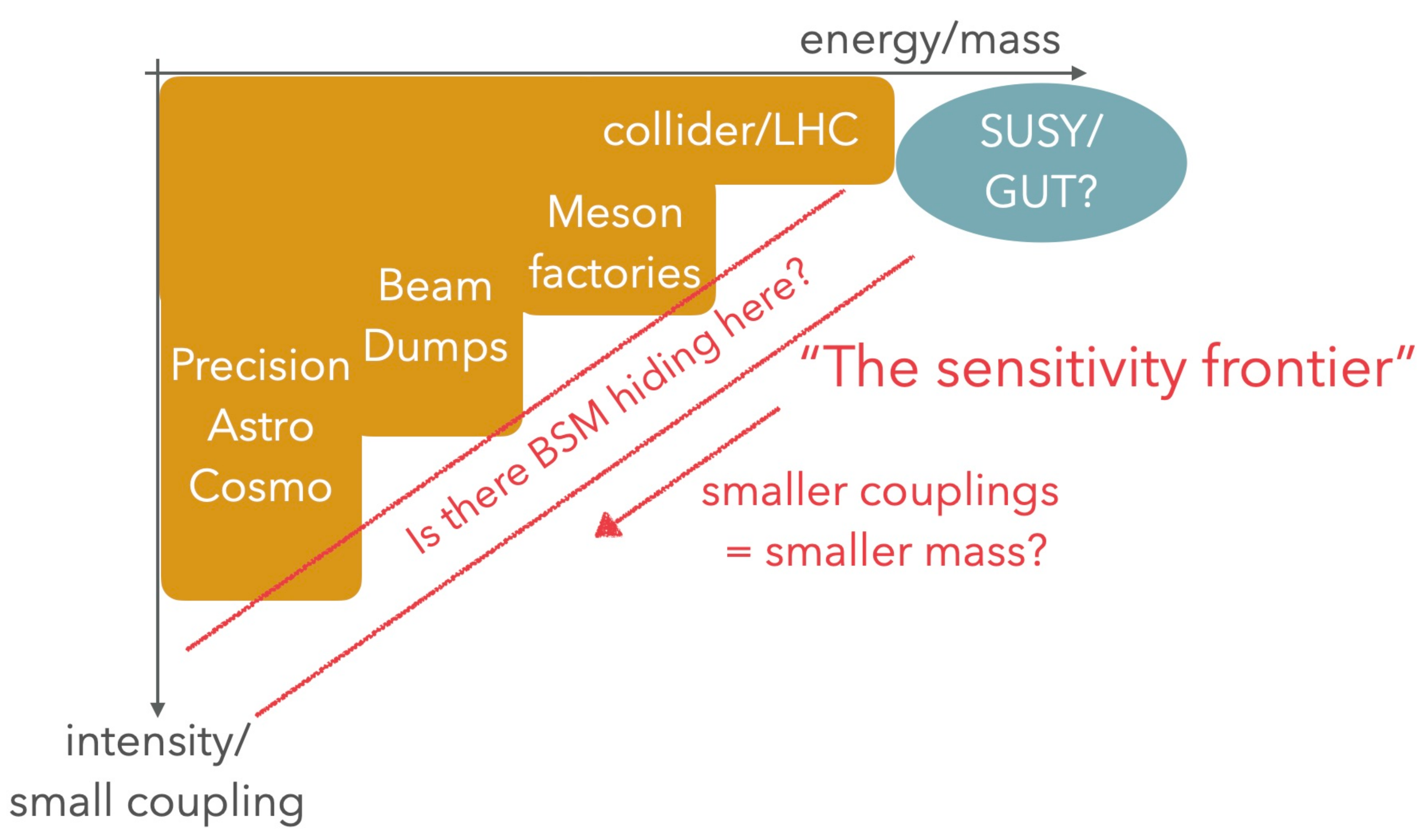}
    \caption{The current sensitivity frontier in the landscape of experimental searches for new physics beyond the Standard Model.}
    \label{fig:sensitivity}
\end{figure}

Experimental evidence like the gravitational observation of dark matter (DM) and the detection of neutrino oscillations have firmly established the existence of new physics beyond the Standard Model (SM).
In the past, these hints have lead many physicists to construct theories of new physics completing the SM at high energy scales, like for example supersymmetric theories, models of grand unification or string theory. A typical shared characteristic  of such ultra-violet (UV) completions is the presence of novel heavy states that can, in principle, couple sizeably to the SM sector. 
Such new heavy states can be tested for example at high-energy experiments like particle colliders. 
However, as illustrated in~\cref{fig:sensitivity} the landscape of particle physics experiments is much more diverse with a plethora of observational strategies testing physics at low energies with ever increasing intensities. Among these are meson factories and beam dump experiments, or astrophysical and cosmological probes. In general, new gauge bosons of an extra $U(1)_X$ symmetry are well-motivated candidates for novel particles that can naturally have ever smaller masses as their coupling to the SM decreases, i.e.~that live at the \textit{sensitivity frontier} of the experimental landscape.

In the minimal hidden photon scenario the gauge boson associated to an additional $U(1)_{X}$ symmetry is kinetically mixed with the SM photon via the operator~\cite{Okun:1982xi,Holdom:1985ag}
\begin{equation}\label{eq:a_mix}
    \mathcal{L} \supset - \frac{\epsilon_A}{2} F_{\mu\nu} X^{\mu\nu}\,,
\end{equation}
where $F_{\mu\nu}$ and $X_{\mu\nu}$ denote the $U(1)_\mathrm{em}$ and $U(1)_{X}$ field strength tensors, respectively. 
Since this term is a a gauge-invariant, renormalisable operator, the kinetic mixing parameter $\epsilon_A$, in principle, is a free parameter of the theory. 
However, in many non-minimal hidden photon models $\epsilon_A$ is generated at the loop level via vacuum polarisation diagrams as the one shown in the right panel of~\cref{fig:b_mixing} due to fermions charged under both $U(1)$ symmetries running in the loop. In these models the loop-induced kinetic mixing typically scales as $\epsilon_A\propto g_x/16\pi^2$, with $g_x$ denoting the coupling constant of $U(1)_{X}$.

The kinetic mixing term in~\cref{eq:a_mix} can be diagonalised by a non-unitary field transformation of the kind
\begin{align}
    A^\mu \to A^\mu - \epsilon_A \, X^\mu \qquad \Rightarrow \qquad  e\,  A_\mu\,  j^\mu_\mathrm{em} \to e\,  A_\mu\,  j^\mu_\mathrm{em} - \epsilon_A e\,  X_\mu\,  j^\mu_\mathrm{em}\,. 
\end{align}
This field redefinition induces a coupling of the new $X$ boson to the SM electromagnetic current $j^\mu_\mathrm{em}$. This interaction motivates the name \textit{hidden photon} for the $X$ boson, since it couples to the QED current analogously to the SM photon, but  suppressed by $\epsilon_A$.

This new hidden photon can generically acquire mass. In the most simple case, the novel $U(1)_X$ symmetry is Higgsed, \textit{i.e.}~it is broken by the vacuum expectation value (VEV) $f$ of a new scalar singlet $S$,
\begin{align}
    \mathcal{L} = (D_\mu S)^\dagger D^\mu S \supset \frac{g_x^2 f^2}{2} \, X_\mu X^\mu \,.
\end{align}
Hence, the mass of the hidden photon, $m_{A'}\propto g_x\, f$, is proportional to the $U(1)_X$ coupling $g_x$. Thus, the smaller the gauge coupling $g_x$ (or the feebler the interactions of the hidden photon) the smaller the mass of the hidden photon. This mechanism makes hidden photons a prime candidate for new physics hiding along the sensitivity frontier illustrated in~\cref{fig:sensitivity} and warrants for a careful study of matching a potential UV hidden photon model onto the low-energy QED regime.

%%%%%%%%%%%%%%%%%%%%%%%%%%%%%%%%%%%%%%%%%%%%%%%
\section{A closer look at the origin of kinetic mixing}
\label{sec:consistent_mix}
%%%%%%%%%%%%%%%%%%%%%%%%%%%%%%%%%%%%%%%%%%%%%%%

\begin{figure}
    \centering
    \includegraphics[width=.3\textwidth] {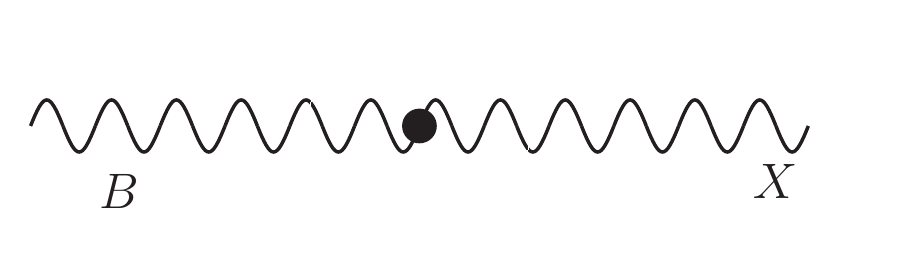}%
    \includegraphics[width=.3\textwidth]{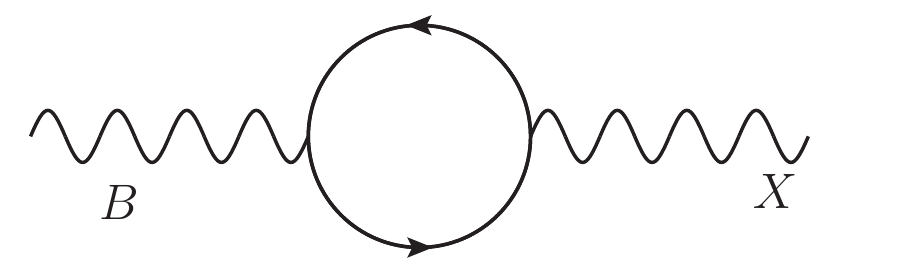}
    \caption{Diagrams of kinetic mixing between the hypercharge boson $B_\mu$ and the $U(1)_{X}$ boson $X_\mu$ at tree level (left) and one-loop level (right).}
    \label{fig:b_mixing}
\end{figure}

The kinetic mixing in~\cref{eq:a_mix} of the $U(1)_X$ boson with the photon of QED cannot be fundamental as the $U(1)_\mathrm{em}$ only arises after electroweak symmetry breaking (EWSB). Hence, we want to study how this operator arises from mixing in the underlying UV theory in the unbroken phase.

%==============================================
\paragraph{Naive picture.}
%==============================================

In the literature it is often assumed that the fundamental mixing of the hidden photon is not with the SM photon, but with the hypercharge boson $B$ of $U(1)_Y$,
\begin{equation}\label{eq:b_mix}
    \mathcal{L} \supset - \frac{\epsilon_B}{2} B_{\mu\nu} X^{\mu\nu}\,,
\end{equation}
where $B_{\mu\nu}$ denotes  the field strength tensor of the hypercharge boson.
This mixing term can  be either elementary or generated at the loop level through fermions carrying charge under both $U(1)_Y$ and $U(1)_{X}$. These two cases are illustrated by the diagrams in~\cref{fig:b_mixing}.
After decomposing the hypercharge boson into its mass eigenstate components, $B_\mu = c_w A_\mu - s_w Z_\mu$, where $c_w\equiv\cos\theta_W$ and $s_w\equiv\sin\theta_W$  denote the cosine and sine of the Weinberg angle $\theta_W$, the mixing term in~\cref{eq:b_mix} reads
\begin{equation}\label{eq:b_dec_mix}
    \mathcal{L} \supset - c_w\frac{\epsilon_B}{2} F_{\mu\nu} X^{\mu\nu}  + s_w\frac{\epsilon_B}{2} Z_{\mu\nu} X^{\mu\nu} \,.
\end{equation}
Matching the terms in~\cref{eq:a_mix} and~\cref{eq:b_dec_mix}, we find the simple expression
\begin{equation}\label{eq:naive_mix}
    \epsilon_A = c_w\, \epsilon_B\,,
\end{equation}
relating the fundamental mixing of the hidden photon with the hypercharge boson, $\epsilon_B$, and
the mixing with the SM photon in the broken phase, $\epsilon_A$.

%==============================================
\paragraph{The full picture.}
%==============================================

A more careful treatment of the matching procedure reveals that the above matching relation in~\cref{eq:naive_mix} cannot be the full picture. In fact, there
exists a dimension-six operator inducing mixing between the $U(1)_X$ and the $SU(2)_L$ bosons~\cite{Bauer:2022nwt},
\begin{equation}\label{eq:d6_op}
    \mathcal{O}_{WX} = \frac{c_{WX}}{\Lambda^2}\, H^\dagger \sigma^i H \, W^i_{\mu\nu} X^{\mu\nu}\,.
\end{equation}
Here $H$ denotes the SM Higgs doublet, $W^i_{\mu\nu}$ is the $SU(2)_L$ field strength tensor, and $\Lambda$ represents the scale of new physics at which this operator is generated, e.g.~by integrating out some heavy new fields.
In the broken phase this operator leads to an effective kinetic mixing term between the neutral component of the weak bosons, $W^3$, and the hidden photon of the form
\begin{equation}\label{eq:wx_mix}
    \mathcal{O}_{WX} \supset - \frac{\epsilon_W}{2} \, W^3_{\mu\nu} X^{\mu\nu}\,,
\end{equation}
where we have identified $\epsilon_W \equiv c_{WX}\, v^2/\Lambda^2$ with the Higgs VEV $v$.
In analogy to what we did above, we also decompose the neutral weak boson $W^3$ into its  mass eigenstate components, $W^3_\mu = s_w A_\mu + c_w Z_\mu$, which leads  to a kinetic mixing term of,
\begin{equation}
    \mathcal{O}_{WX} \supset - s_w \frac{\epsilon_W}{2} \, F_{\mu\nu} X^{\mu\nu} - c_w \frac{\epsilon_W}{2} \, Z_{\mu\nu} X^{\mu\nu}\,.
\end{equation}
Combining this with~\cref{eq:b_dec_mix}, our matching relation~\cref{eq:naive_mix} is modified to also account for the  mixing contribution with the weak boson $W^3$,
\begin{equation}\label{eq:full_mix}
    \epsilon_A = c_w\, \epsilon_B + s_w \, \epsilon_W\,.
\end{equation}

This is a very important result, since in generic hidden photon models with $SU(2)_L$ multiplets charged under the novel $U(1)_{X}$, the operator~\cref{eq:d6_op} will necessarily be generated at the one-loop level.
In these models, the loop contribution to the $W^3-X$ mixing in~\cref{eq:wx_mix} can be computed in analogy to the standard Abelian mixing case~\cite{Bauer:2022nwt}. We identify the kinetic mixing contribution as the transverse component $\Pi_{WX}$ of the full vacuum polarisation amplitude,
\begin{align}
     \Pi^{\mu\nu}_{WX} & = \Pi_{WX}\, [g^{\mu\nu} p_1\cdot p_2  - p_1^\mu p_2^\nu] + \Delta_{WX} \,  g^{\mu\nu} \,,
\end{align}
The loop contribution to the kinetic mixing is then computed as
\begin{equation}
    {\Pi_{WX} \!=\! - \frac{ g\,g_x}{8\pi^2}\,{\sum_f} \int^1_0\!\! dx \,x(1-x) \,{T^f_3} \,{\big(v^f_{X} + a^f_{X}\big) }\,  {\log\left(\frac{\mu^2}{m_f^2-x(1-x)q^2}\right)}} \,,
\end{equation}
where the sum includes all $SU(2)_L$ degrees of freedom $f$ with weak charge $T^f_3$ also  charged under $U(1)_{X}$ with vector and axial-vector couplings $v^f_X$ and $a^f_X$, respectively.

%%%%%%%%%%%%%%%%%%%%%%%%%%%%%%%%%%%%%%%%%%%%%%%
\section{A concrete example: kinetic mixing in $U(1)_{L_\mu-L_\tau}$}
%%%%%%%%%%%%%%%%%%%%%%%%%%%%%%%%%%%%%%%%%%%%%%%

In generic $U(1)_{X}$ models, SM fermions can also be charged under the new symmetry, leading to a gauge interaction of the hidden photon of the type,
\begin{equation}
    \mathcal{L}_\mathrm{int} = - g_x \, j_X^\mu X_\mu\,,
\end{equation}
where \textit{a priori} the current $j_X^\mu = \sum_\psi q_\psi\, \bar \psi \gamma^\mu \psi$ 
can include all SM matter fields, especially also the $SU(2)_L$ quark and lepton doublets $\psi= Q, L$. 
Restricting the gauge current $j_X^\mu$ to only contain SM fields (\textit{i.e.}~disallowing for any new fermions), the minimally anomaly-free models are $U(1)_{B-L}, U(1)_{L_\mu-L_e}, U(1)_{L_e-L_\tau}, U(1)_{L_\mu-L_\tau}$ and linear combinations of these.
In these models, at the very least, (some of) the lepton doublets $L_i$ are charged under the new $U(1)$ symmetry such that $\mathcal{O}_{WX}$ is induced via loops at the renormalizable level (since the scale at which this operator is generated is the electroweak scale, $\Lambda = v$). We will now study how such a loop-generated term affects kinetic mixing in the electroweak-broken and -symmetric phase in the  example of $U(1)_{L_\mu-L_\tau}$.

In the broken phase we can perform the usual, well-known QED mixing computation with two Dirac fermions $f=\mu,\tau$ running in the loop. In the infrared (IR) limit of zero momentum transfer, $q=0$, the resulting mixing parameter reads
\begin{align}\label{eq:photon_mix}
    \epsilon_A = \frac{e\, g_{\mu\tau}}{6 \pi^2} \, \log\left(\frac{m_\mu}{m_\tau}\right)\,, &&  \qquad \qquad \qquad \qquad
    \vcenter{\hbox{\includegraphics[width=.35\textwidth]{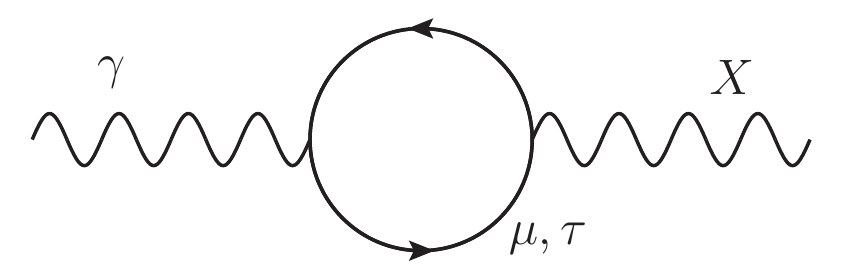}}} \,.
\end{align}
Simultaneously, the \textit{naive} UV computation, in which we only account for mixing with the hypercharge boson $B$, results in a mixing coefficient of 
\begin{align}\label{eq:hyper_mix}
    \epsilon_B = \frac{g'\, g_{\mu\tau}}{24 \pi^2} \, \left[3\log\left(\frac{m_\mu}{m_\tau}\right) + \log\left(\frac{m_{\nu_\mu}}{m_{\nu_\tau}}\right)\right]\,, &&
    \vcenter{\hbox{\includegraphics[width=.35\textwidth]{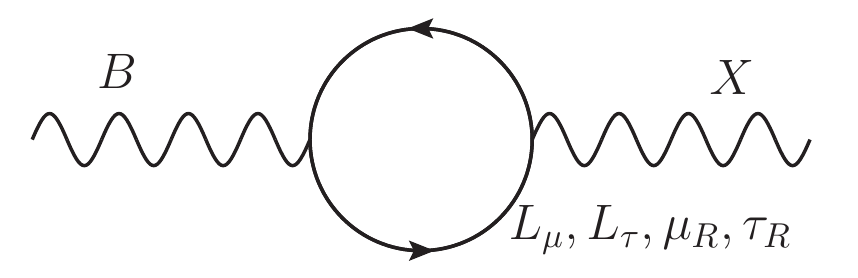}}} \,.
\end{align}
Obtaining these two results, \cref{eq:photon_mix} and ~\cref{eq:hyper_mix}, we have explicitly confirmed that the naive matching relation in~\cref{eq:naive_mix} does not hold and therefore cannot be the correct prescription. 
From our considerations in~\cref{sec:consistent_mix} we already know how to amend the naive mixing prescription, such that the computations of the mixing in the broken and unbroken phase match. 
The solution is to also take into account the loop-induced mixing between the hidden photon and the neutral $SU(2)_L$ boson.
In $U(1)_{L_\mu-L_\tau}$ the second and third generation lepton carry charge under the new symmetry. Hence, a $W^3-X$ mixing term  is generated from the mixing with the diagram with $L_\mu$ and $L_\tau$ running in the loop,
\begin{align}
    \epsilon_W = \frac{g\, g_{\mu\tau}}{24 \pi^2} \, \left[\log\left(\frac{m_\mu}{m_\tau}\right) -\log\left(\frac{m_{\nu_\mu}}{m_{\nu_\tau}}\right)\right]\,, &&
    \vcenter{\hbox{\includegraphics[width=.35\textwidth]{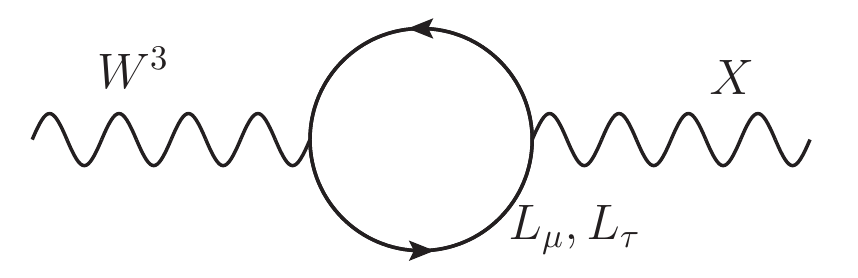}}} \,.
\end{align}
The resulting mixing contribution  exactly yields the missing piece to obtain the mixing coefficient  $\epsilon_A$ of~\cref{eq:photon_mix} in the broken phase  according to the full matching prescription in~\cref{eq:full_mix}. It is particularly noteworthy that the contributions from the neutrinos to $\epsilon_B$ and $\epsilon_W$ exactly cancel. This is expected, since only electrically charged particles can run in the $\gamma-X$ loop, such that neutrinos can never contribute to the hidden photon mixing with the SM QED photon in the broken phase.

%%%%%%%%%%%%%%%%%%%%%%%%%%%%%%%%%%%%%%%%%%%%%%%
\section{The Higgs low-energy theorems}
%%%%%%%%%%%%%%%%%%%%%%%%%%%%%%%%%%%%%%%%%%%%%%%

As a byproduct of computing all the neutral boson vacuum polarisation amplitudes, $\Pi^{\mu\nu}_{V_i V_j}$, we can derive universal expressions for the decay amplitudes of the Higgs to a pair of neutral bosons from the low-energy theorem~\cite{Ellis:1975ap,Shifman:1979eb},
\begin{equation}
    \lim_{p_h\to0} \mathcal{M}(h\to V_i V_j) \to  \frac{\partial}{\partial v} \mathcal{M}(V_i\to V_j).
\end{equation}

Starting from the general one-loop corrected effective low-energy Lagrangian in the electroweak broken phase,
the different mixing contributions can be written as
\begin{align}
    \mathcal{L} = - \frac{1}{4}\,
    (F_{\mu\nu}, Z_{\mu\nu}, X_{\mu\nu})
   \left[\begin{pmatrix}
    1 & 0 & \epsilon_A \\
    0 & 1 & \epsilon_Z \\
    \epsilon_A & \epsilon_Z & 1
    \end{pmatrix}
    +
    \boldsymbol{\Pi}
    \right]
    \begin{pmatrix}
    F^{\mu\nu}\\
    Z^{\mu\nu}\\
    X^{\mu\nu}
    \end{pmatrix}  +  \frac{1}{2} \, 
    (A_{\mu},Z_\mu, X_{\mu})\,
    \left[  \mathbf{M}+\mathbf{\Delta} \right]
    \begin{pmatrix}
    A^{\mu}\\
    Z^\mu \\
    X^{\mu}
    \end{pmatrix}\,, \label{eq:loop_lag}
\end{align}
where $\boldsymbol{M}=\text{diag}(0,m_Z^2,m_X^2)$ denotes the tree-level mass matrix of the neutral bosons and $\epsilon_A$ and $\epsilon_Z$ are the tree-level kinetic mixing coefficients. Furthermore, the loop-generated contributions to kinetic and mass mixing are encoded in the matrices
\begin{align}\label{eq:loop_matr}
    \boldsymbol{\Pi}=
    \begin{pmatrix}
    \Pi_{\gamma\gamma} & \Pi_{\gamma Z} & \Pi_{\gamma X} \\
    \Pi_{\gamma Z} & \Pi_{ZZ} & \Pi_{ZX} \\
    \Pi_{\gamma X} & \Pi_{ZX} & \Pi_{XX}
    \end{pmatrix}
    \,, &&
    \boldsymbol{\Delta} = \begin{pmatrix}
    0&0 & 0 \\
    0& \Delta_{ZZ} & \Delta_{ZX} \\
    0& \Delta_{ZX} & \Delta_{XX}
    \end{pmatrix}\,.
\end{align}
We can diagonalise the tree-level kinetic mixing terms in~\cref{eq:loop_lag} via a non-unitary field redefinition given by 
\begin{equation}
  G = 
\begin{pmatrix}
 1 & 0 & -{\frac{\epsilon _A}{ \sqrt{1-\epsilon _A^2-\epsilon _Z^2}}} \\[8pt]
 0 & 1 & -{\frac{\epsilon _Z}{ \sqrt{1-\epsilon _A^2-\epsilon _Z^2}}} \\[8pt]
 0 & 0 & {\frac{1}{ \sqrt{1-\epsilon _A^2-\epsilon _Z^2}}} 
\end{pmatrix}\,.
\end{equation}
After diagonalisation we find the general Higgs decay amplitude to read~\cite{Bauer:2022nwt},
\begin{align}
     \mathcal{M}_{h\to V_i\,V_j}^{\mu\nu} =\, {\partial_v} [  G^T\,\boldsymbol{\Pi} \,G]_{ij}\ [p_2^\mu\, p_1^\nu -  p_1\cdot p_2\, g^{\mu\nu}] \ + \  {\partial_v} \big[G^T\,\big[\boldsymbol{M}+\boldsymbol{\Delta}\big] \,G \big]_{ij} \  g^{\mu\nu} \,,
    \label{eq:low_en_amp}
\end{align}
where we have factored out the gauge boson polarisation vectors of the amplitude $\mathcal{M}_{h\to V_i\,V_j} = \mathcal{M}_{h\to V_i\,V_j}^{\mu\nu}\ \epsilon^*_{\mu,\lambda}(p_1)\,\epsilon^*_{\nu,\lambda'}(p_2)$. To leading order in the small mixing coefficients $\epsilon_A$ and $\epsilon_Z$, for the rotated matrices in~\cref{eq:loop_matr}  we find the symmetric matrices,
\begin{align}\label{eq:vac_pol_rot}
         G^T\,\boldsymbol{\Pi} \,G &=\boldsymbol{\Pi} -
    \begin{pmatrix}
    0 & 0 &   \epsilon_A\,\Pi_{\gamma\gamma}+ \epsilon_Z\,\Pi_{\gamma Z} \\[.25cm]
    \cdot &0  &   \epsilon_A\,\Pi_{\gamma Z}+ \epsilon_Z\,\Pi_{ZZ} \\[.25cm]
    \cdot & \cdot & 2\epsilon_A \Pi_{\gamma X} + 2\epsilon_Z \Pi_{Z X}
    \end{pmatrix}\,, \\
    G^T\,\big[\boldsymbol{M}+\boldsymbol{\Delta}\big] \,G  &=\big[\boldsymbol{M}+\boldsymbol{\Delta}\big]-
    \begin{pmatrix}
    0 & 0 &   0\\
    \cdot & 0 &    \epsilon_Z\,(m_Z^2+\Delta_{ZZ}) \\[.25cm]
    \cdot & \cdot &  2\epsilon_Z\, \Delta_{Z X} 
    \end{pmatrix}\,,
\end{align}
Note that the mass mixing terms, $\Delta_{V_i V_j}$, are only generated in theories where the loop fermions have axial-vector charges under both gauge groups.

In the case that only SM fermions are charged under the novel $U(1)_X$ symmetry, the expressions relevant for computing the Higgs decay amplitudes to photons, Z and X bosons according to~\cref{eq:low_en_amp} are given by
\begin{align} \label{eq:1loop_GX}
    {\partial_v}\, \Pi_{\gamma X}(0) &= \sum_f N_c^f \frac{e\,g_x}{12\,\pi^2 \,v} \,Q_f\,v^f_{X} \,, \\
    {\partial_v}\, \Pi_{Z X}(0) &= \sum_f N_c^f \frac{e\,g_x}{24\,\pi^2 \,v} \,\frac{T_3^f - 2 \,s_w^2\,Q_f}{s_w c_w}\,v^f_{X} \,, \\
    {\partial_v}\, \Pi_{X X}(0) &= \sum_f N_c^f \frac{g_x^2}{24\,\pi^2\, v} \, 
    v^{f2}_{X} \,, \label{eq:1loop_XX}
\end{align}
where the sum runs over all heavy fermions with with $m_f \gg m_h$, \textit{i.e.}~only the top quark in the SM.

\begin{figure*}
    \centering
    \includegraphics[width=.95\textwidth]{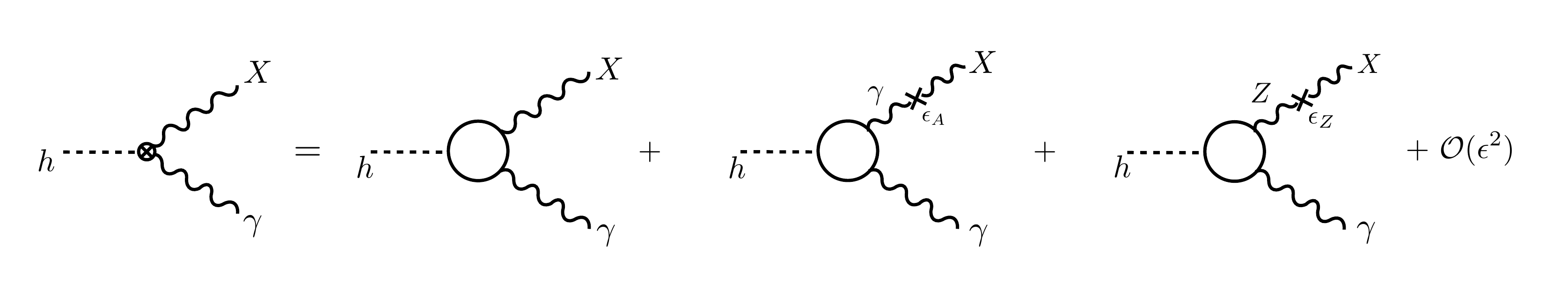}
    \caption{Diagrammatic representation of the amplitude $\mathcal{M}(h \to \gamma X)$ due to the various contributions from fermion loops at linear order in the kinetic mixing parameters $\epsilon_A$ and $\epsilon_Z$.}
    \label{fig:diagrams}
\end{figure*}

In a practical example of the Higgs low-energy theorems, we can compute the branching ratios of the SM Higgs decaying to $\gamma X$ and $XX$ in a model of gauged $U(1)_{B-L}$. For example, the relevant contributions to the decay $h\to \gamma X$ are shown in a diagrammatic representation in~\cref{fig:diagrams} to leading order in the kinetic mixing parameters $\epsilon_A$ and $\epsilon_Z$. Assuming the new gauge boson to be light, $m_X\ll m_h$, we can universally express the branching ratios to leading order in the gauge coupling $g_x$ and the kinetic mixing parameter as
\begin{align}\label{eq:br_GX}
\mathcal{BR}_{h\to \gamma X}\!&\simeq (0.92\, g_x^2 +6.36g_x\epsilon_A  + 11.01\epsilon^2_A)\,\cdot 10^{-3}\!, \\[3pt]
\mathcal{BR}_{h\to X X}  \!&\simeq g_x^2 (2.5\, g_x^2 - 5.7 \, g_x\epsilon_A + 3.2 \epsilon^2_A)\,\cdot 10^{-3}. \label{eq:br_XX}
\end{align}
Evaluating these expressions for still allowed values of the gauge coupling and mixing of $g_x\sim10^{-4}$ and $\epsilon_A\sim10^{-3}$ yields model-independent branching ratios of $\mathcal{BR}_{h\to \gamma X} \sim 10^{-8}$ and $\mathcal{BR}_{h\to X X} \sim 10^{-17}$.
While the process $h\to XX$ seems hopeless to be tested at any conceivable future detector, the process $h\to\gamma X$ could be tested at an upcoming collider like the FCC-hh aiming at collecting up to $\mathcal{O}(10^{10})$ Higgs bosons.

%%%%%%%%%%%%%%%%%%%%%%%%%%%%%%%%%%%%%%%%%%%%%%%
\section{Conclusion}
%%%%%%%%%%%%%%%%%%%%%%%%%%%%%%%%%%%%%%%%%%%%%%%

In summary, hidden photons are well-motivated candidates for new physics hiding along the experimental \textit{sensitivity frontier}. 
In the minimal setup, the interactions of these particles with the SM sector arises purely through kinetic mixing.
Due to gauge invariance the kinetic mixing of the novel $X$ boson has to proceed with the hypercharge boson $B$ in the electroweak symmetric phase.
At dimension-six, however, there exists an operator coupling the $X$ boson to the $SU(2)_L$ bosons, generating a mixing term between the hidden photon and $W^3$, which can effectively arise at the renormalisable level.
In theories in which $SU(2)_L$ multiplets are carrying charge under the new $U(1)_X$ symmetry this novel type of mixing is always generated at the one-loop level. It is vital to take this $W^3-X$ mixing into account in order to obtain the correct matching onto the effective mixing with the photon in the electroweak broken phase.
In essence, the correct matching of the mixing of the hidden photon with the hypercharge and neutral weak boson, $\epsilon_B$ and $\epsilon_W$, onto the mixing with the photon, $\epsilon_A$, is given by the relation 
\begin{equation*}
    {\epsilon_A = c_w\, \epsilon_B + s_w \, \epsilon_W}\,.
\end{equation*}
Importantly, the weak mixing contribution $\epsilon_W$ is unavoidably generated at the one-loop level in the phenomenologically interesting anomaly-free hidden photon models like $U(1)_{B-L}$, $U(1)_{L_\mu-L_e}$, $U(1)_{L_e-L_\tau}$, $U(1)_{L_\mu-L_\tau}$, and combinations of these.

Finally, we have demonstrated how to obtain the decay amplitudes of the Higgs to a pair of neutral bosons from the vacuum polarisation amplitudes via the Higgs low-energy theorems. This method automatically generates all relevant contributions to the decay amplitude at a fixed order in the kinetic mixing.

%%%%%%%%%%%%%%%%%%%%%%%%%%%%%%%%%%%%%%%%%%%%%%%
\section*{Acknowledgements}
%%%%%%%%%%%%%%%%%%%%%%%%%%%%%%%%%%%%%%%%%%%%%%%
\noindent
I would like to thank my collaborator Martin Bauer for the fruitful collaboration that has lead to this work. PF is supported by the Spanish Agencia Estatal de Investigaci\'on through the grants PID2021-125331NB-I00 and CEX2020-001007-S, funded by 
MCIN/AEI/10.13039/501100011033.

%==============================================
% REFERENCES
%==============================================

\bibliographystyle{JHEP}
\bibliography{literature}

\end{document}